\documentclass[article,aps,nofootinbib,showpacs,amsfonts,epsf]{revtex4}

\input{epsf.tex}

\newcommand{\sss}{/\hspace{-0.45em}s}
\newcommand{\E}{{\cal{E}}}

\newcommand{\s}{\sigma}

\renewcommand{\a}{\alpha}
\renewcommand{\k}{\kappa}

\newcommand{\be}{\begin{equation}}
\newcommand{\ee}{\end{equation}}
\newcommand{\bea}{\begin{eqnarray}}
\newcommand{\eea}{\end{eqnarray}}
\newcommand{\ba}{\begin{array}}
\newcommand{\ea}{\end{array}}
\def\J#1#2#3#4{{#1} {\bf #2}, #3 (#4)}
\def\PRD{Phys. Rev. D}
\def\PR{Phys. Rev.}
\def\PRL{Phys. Rev. Lett.}
\def\PTP{Prog. Theor. Phys.}
\def\APN{Ann. Phys. (NY)}
\def\APL{Ann. Phys. (Leipzig)}

\def\JMP{J. Math. Phys.}
\def\CPAM{Comm. Pure Appl. Math.}

\def\CQG{Class. Quantum Grav.}

\def\PLA{Phys. Lett. A}

\begin{document}
\draft

\title{Stationary configurations of two extreme black holes\\ obtainable from the Kinnersley-Chitre solution}

\author{V.~S.~Manko,$^*$ E.~Ruiz$^\dag$ and M. B. Sadovnikova$^\ddag$}
\address{$^*$Departamento de F\'\i sica, Centro de Investigaci\'on y de
Estudios Avanzados del IPN, A.P. 14-740, 07000 M\'exico D.F.,
Mexico\\$^\dag$Instituto Universitario de F\'{i}sica
Fundamental y Matem\'aticas, Universidad de Salamanca, 37008 Salamanca, Spain\\$^\ddag$Department of Quantum Statistics and Field Theory, Lomonosov Moscow State University, Moscow 119899, Russia}

\begin{abstract}
Stationary axisymmetric systems of two extreme Kerr sources separated by a massless strut, which arise as subfamilies of the well-known Kinnersley-Chitre solution, are studied. We present explicit analytical formulas for the individual masses and angular momenta of the constituents and establish the range of the parameters for which such systems can be regarded as describing black holes. The mass-angular momentum relations and the interaction force in the black-hole configurations are also analyzed. Furthermore, we construct a charging generalization of the Kinnersley-Chitre metric and, as applications of the general formulas obtained, discuss two special cases describing a pair of identical co- and counterrotating extreme Kerr-Newman black holes kept apart by a conical singularity. From our analysis it follows in particular that the equality $m^2-a^2-e^2=0$ relating the mass, angular momentum per unit mass and electric charge of a single Kerr-Newman extreme black hole is no longer verified by the analogous extreme black-hole constituents in binary configurations.  \end{abstract}

\pacs{04.20.Jb, 04.70.Bw, 97.60.Lf}

\maketitle


\section{Introduction}

The well-known Kinnersley-Chitre (KC) solution \cite{KCh} represents the extremal limit of the double-Kerr spacetime \cite{KNe}, and as such permits one to describe stationary axisymmetric configurations of two extreme Kerr sources. In the case of the balancing constituents \cite{Tom,DHo,CMR} it can be shown that at least one of these constituents is endowed with a negative mass, which spoils the interpretation of those binary configurations as describing two black holes. This naturally motivates the search for (and subsequent analysis of) the systems composed of two extreme black holes separated by a massless strut (conical singularity \cite{Isr}) which may arise for instance in the context of some more general non-stationary axisymmetric scenarios as {\it momentary stationary data} \cite{DOr} for interacting black holes. In the recent paper \cite{MRu}, various binary configurations of extreme Kerr sources with a middle strut were obtained in the analytical form thanks to a new simple representation of the KC metric, and some examples were given in which both extreme constituents can be envisaged as black holes due to the positiveness of their Komar \cite{Kom} masses and regularity of the corresponding spacetimes outside the horizons.

The present paper pursues the following two main goals. First, we would like to amplify the initial analysis of the novel binary configurations discovered in \cite{MRu}, with the general analytic formulas for the individual Komar quantities (masses and angular momenta) characterizing the extreme constituents in the case of {\it all four} subfamilies of the KC solution from \cite{MRu} whose total mass can take positive values. This will allow us to make a systematic study of the mass-angular momentum relations in the corresponding purely black-hole binary systems. Second, by using the Harrison-Ernst invariance transformation \cite{Har,Ern}, we shall construct a charging generalization of KC metric, which will permit us to get the first known configurations of two extreme Kerr-Newman (KN) constituents \cite{New} kept apart by a massless strut, and study in more detail two particular, equatorially symmetric and antisymmetric, cases.

The plan of the paper is as follows. In Sec.~II we revise the KC metric and analyze the mass-angular momentum relations and the interaction force in the configurations composed of two extreme Kerr black holes. Here we present the concise analytical formulas for the individual Komar characteristics of the constituents and plot the stationary limit surfaces (SLS) for various particular configurations. In Sec.~III a charging generalization of KC metric is constructed and certain electrovac solutions describing two extreme KN constituents separated by a conical singularity are identified. The particular cases of two identical co- and counterrotating KN extreme black holes are analyzed in more detail. The discussion of the results obtained and concluding remarks can be found in Sec.~IV.

\section{The KC metric and configurations of two extreme Kerr black holes}

The five-parameter KC solution, generated in the paper \cite{KCh} with the aid of a special subgroup of the KC symmetry transformations preserving asymptotic flatness, is defined by the complex potential $\E$ of the form
\bea
\E&=&(A-B)/(A+B), \nonumber\\
A&=&p^2(x^4-1)+(\a^2-\beta^2)(x^2-y^2)^2+q^2(y^4-1) -2ipqxy(x^2-y^2) \nonumber\\
&-&2i\a(x^2+y^2-2x^2y^2)-2i\beta xy(x^2+y^2-2), \nonumber\\
B&=&2(P-iQ)[px(x^2-1)+iqy(y^2-1)-i(p\a+iq\beta)x(x^2-y^2) \nonumber\\ &+&i(p\beta+iq\a)y(x^2-y^2)], \label{KC_sol} \eea
which satisfies the Ernst equation \cite{Ern2}:
\be
{\rm Re}(\E)\left\{\frac{\partial}{\partial x}\left[(x^2-1)\frac{\partial\E}{\partial x}\right] + \frac{\partial}{\partial y}\left[(1-y^2)\frac{\partial\E}{\partial y}\right]\right\} = (x^2-1)\left(\frac{\partial\E}{\partial x}\right)^2 +(1-y^2)\left(\frac{\partial\E}{\partial y}\right)^2. \label{E_eq} \ee
In the above equations, $x$ and $y$ are prolate spheroidal coordinates related to the usual Weyl-Papapetrou cylindrical coordinates $\rho$ and $z$ via the formulas
\be
x=\frac{1}{2\k}(r_++r_-), \quad y=\frac{1}{2\k}(r_+-r_-), \quad r_\pm=\sqrt{\rho^2+(z\pm\k)^2}, \label{xy} \ee
and $\kappa$, $\a$, $\beta$, $p$, $q$, $P$, $Q$ are real parameters, the latter four of which are subjected to the constraints
\be
p^2+q^2=1, \quad P^2+Q^2=1. \label{pP} \ee

The Papapetrou stationary axisymmetric line element \cite{Pap} in the coordinates $x$, $y$ takes the form
\be
d s^2=\kappa^2f^{-1}\left[e^{2\gamma}(x^2-y^2)\left(\frac{d x^2}{x^2-1}+\frac{d y^2}{1-y^2}\right)+(x^2-1)(1-y^2)d\varphi^2\right]-f(d
t-\omega d\varphi)^2, \label{Pap} \ee
and concise expressions for the metric functions $f$, $\gamma$, $\omega$ of KC solution have been obtained in \cite{MRu} by making use of Perj\'es' representation \cite{Per} of the Tomimatsu-Sato metrics \cite{TSa}\footnote{Mention that the study of the factor structure of the Tomimatsu-Sato spacetimes was initiated by Ernst \cite{Ern3} and Hoenselaers \cite{HEr}.} and also of some results of the paper \cite{Yam}:
\bea f&=&\frac{N}{D}, \quad e^{2\gamma}=\frac{N}{K_0^2(x^2-y^2)^4}, \quad \omega=2J_0 y -\frac{\k(1-y^2)F}{N}, \nonumber\\
N&=&\mu^2-(x^2-1)(1-y^2)\s^2, \nonumber\\ D&=&N+\mu\pi+(1-y^2)\s\tau, \nonumber\\ F&=&(x^2-1)\s\pi+\mu\tau, \nonumber\\ \mu&=&p^2(x^2-1)^2+q^2(1-y^2)^2+(\a^2-\beta^2)(x^2-y^2)^2, \nonumber\\ \s&=&2[pq(x^2-y^2)+\beta(x^2+y^2)-2\a xy], \nonumber\\ \pi&=&(4/K_0)\{K_0 [pPx(x^2+1)+2x^2+qQy(y^2+1)] \nonumber\\ &+&2(pQ+pP\a+qQ\beta)[pqy(x^2-y^2)+\beta y(x^2+y^2)-2\a xy^2] \nonumber\\ &-&K_0(x^2-y^2)[(pQ\a-qP\beta)x+(qP\a-pQ\beta)y] \nonumber\\ &-&2(q^2\a^2+p^2\beta^2)(x^2-y^2)+4(pq+\beta)(\beta x^2-\a xy)\}, \nonumber\\ \tau&=&(4/K_0)\{K_0 x [(qQ\a+pP\beta)(x^2-y^2)-qP(1-y^2)] \nonumber\\ &+&(pQ+pP\a+qQ\beta)y[(p^2-\a^2+\beta^2)(x^2-y^2)+y^2-1] \nonumber\\ &-&pQK_0y(x^2-1)-2p(q\a^2-q\beta^2-p\beta)(x^2-y^2) \nonumber\\ &-&(pq+\beta)(1-y^2)\}, \nonumber\\ J_0&=&-2\k(pQ+pP\a+qQ\beta)/K_0, \quad K_0=p^2+\a^2-\beta^2. \label{mf_gen} \eea

The KC solution is describing two extreme Kerr sources kept apart by a massless strut when the condition of asymptotic flatness ($J_0=0$) and the axis condition ($\omega(x=1)=0$) are satisfied simultaneously, and in the paper \cite{MRu} those conditions were solved analytically, giving rise to several families of binary systems composed of extreme Kerr constituents. As was also established in \cite{MRu}, some of these configurations may have two constituents with positive Komar masses, thus representing physical interest as describing pairs of interacting extreme black holes. Since the constituents are not identical in general and, therefore, can have non-equal masses and angular momenta, for carrying out the generic physical analysis of the new configurations it is important to have the full set of analytic expressions for the individual Komar quantities. In what follows we shall consider the mass-angular momentum relations in the configurations of two extreme black holes by using the analytic formulas for the Komar quantities that we have been able to work out recently in a concise form by taking appropriate limits in the respective general expressions previously obtained by Tomimatsu \cite{Tom} and Dietz and Hoenselaers \cite{DHo} for the double-Kerr solution.\footnote{Note that there is a misprint in the first line of Eq.~(5.9) of Ref.~\cite{DHo}: the factor $(q_3+\sss_3)$ must be squared.}

The {\it first} subfamily of extreme Kerr constituents found in \cite{MRu} is defined by $\a$ and $\beta$ of the form
\be
\a=-\frac{pQ(pP+q^2)}{p^2-Q^2}, \quad \beta=\frac{pq(pP+Q^2)}{p^2-Q^2}, \label{par_1} \ee
and in this case the total mass $M$ and total angular momentum $J$ are given by the expressions~\cite{MRu}
\be
M=\frac{2\kappa(pP+q^2)}{p^2-q^2}, \quad J=\frac{2\kappa^2q[(1+2pP)^2-(p+P)^2]}{p(p^2-q^2)^2}. \label{MJ_1} \ee

As we already mentioned above, the individual Komar masses and angular momenta of the components are obtainable from the general formulas derived in \cite{Tom,DHo}. Omitting the cumbersome calculational details, we give here the final results for the masses $M_i$ and angular momenta $J_i$ of the subfamily (\ref{par_1}):
\bea
M_1&=&\frac{\kappa[Q(pP+q^2)+q(pP+Q^2)]}{Q(p^2-q^2)}, \quad M_2=\frac{\kappa[Q(pP+q^2)-q(pP+Q^2)]}{Q(p^2-q^2)}, \nonumber\\
J_1&=&\frac{QM_1^2}{p(1-pP+qQ)}, \quad J_2=-\frac{QM_2^2}{p(1-pP-qQ)}, \label{MiJi_1} \eea
where the subscript 1 labels the upper constituent, and 2 the lower one.

From (\ref{MiJi_1}) it follows immediately that the constituents in this subfamily are counterrotating. Indeed, since the factors $(1-pP\pm qQ)$ in the denominators of $J_i$ cannot take negative values, then $J_1$ and $J_2$ can have only opposite signs, thus proving the counterrotation of the sources.

The expressions (\ref{MiJi_1}) for the individual Komar masses permit us to find all parameter ranges at which both masses $M_1$ and $M_2$ take positive values, i.e., when both constituents are extreme black holes. As the case $q=0$ representing identical counterrotating black holes was already considered in the literature \cite{MRRS}, below we shall restrict ourselves to the case of two non-identical extreme black holes only. Then the resolution of the system of inequalities $M_1>0$, $M_2>0$ readily yields the following two sets of the parameters defined by the positive and negative values of $p$:
\be -1<p<-\frac{1}{\sqrt{2}}, \quad p<P<-|q| \label{pos_1.1} \ee
and
\be
\frac{1}{\sqrt{2}}<p<1, \quad |q|<P<p, \label{pos_1.2} \ee
(the corresponding quantities $q$ and $Q$ may have either sign).

The analysis of the mass-angular momentum relations in the binary black-hole configurations described by the parameter sets (\ref{pos_1.1}) and (\ref{pos_1.2}) reveals the following interesting property of these binary systems: {\it the individual Komar quantities of both black holes verify the inequality $|J_i|>M_i^2$ in general, whereas for the total mass and total angular momentum the inequality $|J|<M^2$ always holds.} This is apparently in contrast to the single Kerr extreme geometry characterized by the equality $|J|=M^2$, but, at the same time, in full agreement with the recent numerical study of the interacting black holes carried out in \cite{DOr} and \cite{CHR}. It is worth noting that in the {\it equilibrium} configurations of two extreme Kerr sources recently considered in the paper \cite{CMR}, the individual Komar quantities verify the inequality $|J_i|\le M_i^2$ characterizing a single Kerr black hole \cite{Ker}, but this of course should be attributed to the presence of a negative mass in all such configurations from \cite{CMR}.

Let us also consider the interaction force between two black holes whose form is given by the formula \cite{Isr,Wei}
\be
{\cal F}=\frac{1}{4}(e^{-\gamma_0}-1), \label{force} \ee
where $\gamma_0$ is the value of the metric function $\gamma$ on the part of the symmetry axis separating the constituents, i.e., in our extreme case $\gamma_0=\gamma(x=1)$. From (\ref{force}), (\ref{mf_gen}) and (\ref{par_1}) we then obtain
\be
{\cal F}=\frac{p^2-Q^2}{4Q^2}, \label{force_1} \ee
and one can easily see that the above ${\cal F}$ is positive definite for all $p$ and $P$ from (\ref{pos_1.1}), (\ref{pos_1.2}), which means that the strut preventing the constituents from falling onto each other cannot be removed.

In Fig.~1 we have plotted the stationary limit surfaces (SLS), defined by the equation $f=0$, for two particular configurations of non-identical counterrotating extreme black holes given by
\be
\kappa=2, \quad p=\frac{12}{13}, \quad q=\frac{5}{13}, \quad P=\frac{3}{5}, \quad Q=\frac{4}{5}, \label{pc_1} \ee and
\be \kappa=2, \quad p\simeq-0.995, \quad q=0.1, \quad P=-0.5, \quad Q\simeq0.866. \label{pc_2} \ee
One can see that each SLS consists of two disconnected regions, and no massless ring singularities are being developed off the symmetry axis. It is worth underlying that the black-hole solutions of the subclass (\ref{par_1}) are regular on the interval $x=1$, $-1<y<1$ of the symmetry axis separating the black holes (in spite of the lack of elementary flatness there requiring $\gamma(x=1)=0$), which is of course the property shared by all the black-hole configurations considered in this paper.

The {\it second} family of two extreme Kerr sources separated by a strut is defined by the parameter choice \cite{MRu}
\bea
\a&=&-\frac{Q[q\Delta+pq^2+P(1+p^2)]}{2(p^2-Q^2)}, \quad \beta=\frac{p[P\Delta+q(1+pP+Q^2)]}{2(p^2-Q^2)}, \nonumber\\
\Delta&=&\sqrt{4p^2(1+pP)+q^2(p+P)^2}, \label{par_2} \eea
and the corresponding total mass and total angular momentum of the system have the form
\bea
M&=&\frac{\kappa[q\Delta-p(1+p^2)-q^2P]}{p(p^2-q^2)}, \nonumber\\ J&=&\frac{M^2}{4p^2}[(2p^2-p^3P+q^2Q^2)\Delta+p^2qP(p-P)^2-q(p+P)(p^2-Q^2)], \label{MJ_2} \eea
where we have used a slightly different representation of $J$ than in \cite{MRu}, having in mind the subsequent analysis of the mass-angular momentum relations.

For the individual Komar quantities of the constituents the following expressions have been found:
\bea
M_1&=&\frac{\kappa[(q+pqP-p^2Q)\Delta-(1+pP)(p+p^3+q^2P-pqQ)+pq^3Q]} {2p(1+pP)(p^2-q^2)}, \nonumber\\
M_2&=&\frac{\kappa[(q+pqP+p^2Q)\Delta-(1+pP)(p+p^3+q^2P+pqQ)-pq^3Q]} {2p(1+pP)(p^2-q^2)}, \nonumber\\
J_1&=&\frac{(1+pP+qQ)M_1^2} {2p(p+P)^2}[(1+pP+q^2)\Delta-4pq+pq(p-P)^2], \nonumber\\
J_2&=&\frac{(1+pP-qQ)M_2^2} {2p(p+P)^2}[(1+pP+q^2)\Delta-4pq+pq(p-P)^2], \label{MiJi_2} \eea
and it is trivial to see that the above $J_1$ and $J_2$ cannot have opposite signs, and hence the extreme constituents are corotating.

We are ready now to establish the black-hole sector of the subfamily (\ref{par_2}) defined by the positive values of the masses $M_1$ and $M_2$. From (\ref{MiJi_2}) we find the desired values of the parameters ensuring $M_i>0$:
\be -\frac{1}{\sqrt{2}}<p<0, \quad q>0, \quad -1<P<-p\,\,\cup\,\,q<P<1, \label{pos_2.1} \ee
and
\be
0<p<\frac{1}{\sqrt{2}}, \quad q<0, \quad -1<P<q\,\,\cup\,\,-p<P<1, \label{pos_2.2} \ee
where we have not included the case of identical corotating extreme black holes ($Q=0$) which was already analyzed in some detail in \cite{MRu}. Note that the sign of $Q$ is unimportant either in (\ref{pos_2.1}) or in (\ref{pos_2.2}).

The black-hole configurations of this subfamily provide us with a variety of possible mass-angular momentum relations for the respective total and individual quantities. Thus, the ratio $|J|/M^2$ involving total quantities can take values less, greater or equal to 1. The inequality $|J|/M^2<1$ holds for all the black-hole configurations defined by (\ref{pos_2.1}) with $q<P<1$, and by (\ref{pos_2.2}) with $-1<P<q$, whereas the configurations with $-1<P<-p$ in (\ref{pos_2.1}) and $-p<P<1$ in (\ref{pos_2.2}) can lead to any of the aforementioned three possibilities regarding the ratio $|J|/M^2$. The analysis of the second formula in (\ref{MJ_2}) reveals that, for any values of $p$ and $q$ in (\ref{pos_2.1}), (\ref{pos_2.2}), either of these possibilities is determined by the roots of some quartic equation for $P$ whose explicit form can be omitted here. This can be illustrated by the following example: if one chooses $p=-7/25$, $q=24/25$, then one arrives, through the resolution of the quartic equation, at the value $P_0=(\sqrt{873849}-2043)/2750\simeq-0.403$, such that the configurations with $-1<P<P_0$ are characterized by the inequality $|J|/M^2<1$, for the configurations with $P_0<P<7/25$ one has $|J|/M^2>1$, and the equality $|J|/M^2=1$ is verified at $P=P_0$. The violation of the equality $|J_i|/M_i^2=1$ by the individual black-hole constituents in this example can be easily demonstrated if one chooses for instance $P=-0.3$, $Q\simeq0.954$, thus finding from (\ref{MiJi_2}) $J_1/M_1^2\simeq8.002$, $J_2/M_2^2\simeq0.673$.

Although in the paper \cite{MRu} some typical SLS in the case of corotating (identical) extreme black holes have already been plotted, in Fig.~2 we give two more examples of SLS involving the non-identical constituents; these arise at the parameter choices
\be
\kappa=2, \quad p\simeq-0.3, \quad q=\simeq0.954, \quad P=0.96, \quad Q=0.28, \label{pc_3} \ee and
\be \kappa=2, \quad p=\frac{3}{5}, \quad q=-\frac{4}{5}, \quad P=Q=\frac{1}{\sqrt{2}}. \label{pc_4} \ee
The appearance of a massless ring singularity in Fig.~2(ii) may look very surprising since both constituents have positive Komar masses. However, this phenomenon could probably be attributed to the non-stationary processes that might occur during the formation of a common SLS.

We skip the third subfamily of extreme Kerr sources found in \cite{MRu} as it is characterized by the negative total mass and hence is not of interest for us here. We go directly to the next {\it special subfamily} of extreme configurations defined by
\be
\a=p, \quad \beta=-p/q, \quad P=0, \quad Q=1, \label{par_4} \ee
with the corresponding expressions for the total mass and total angular momentum:
\be
M=\frac{2\kappa q^2}{p^2-q^2}, \quad J=\frac{2\kappa^2q^3}{p(p^2-q^2)^2}. \label{MJ_4} \ee

Although the total mass in (\ref{MJ_4}) takes positive values when $p^2>q^2$, the individual masses of the constituents are given by the expressions
\be
M_1=\frac{\kappa q(1+q)}{p^2-q^2}, \quad M_2=-\frac{\kappa q(1-q)}{p^2-q^2}, \label{Mi_4} \ee
whence it follows that $M_1$ and $M_2$ always have opposite signs and, therefore, the subfamily (\ref{par_4}) cannot describe two extreme black holes. Note that this subfamily represents counterrotating sources since the individual angular momenta $J_i$ have the form
\be
J_1=\frac{M_1^2}{p(1+q)}, \quad J_2=-\frac{M_2^2}{p(1-q)}, \label{Ji_4} \ee
and hence the ratio $J_1/J_2$ is a negative quantity.

The {\it last subclass} of extreme sources obtained in \cite{MRu} is defined by
\be
\a=\frac{p}{2q}(\Delta_0+q), \quad \beta=-p/q, \quad P=0, \quad Q=1, \quad \Delta_0=\sqrt{4+q^2},  \label{par_5} \ee
and the corresponding total mass and total angular momentum are
\be
M=\frac{\kappa(q\Delta_0-1-p^2)}{p^2-q^2}, \quad J=\frac{[(1+p^2)\Delta_0+q^3]M^2}{4p}. \label{MJ_5} \ee

For the individual Komar quantities of this subfamily we have found the following expressions:
\bea
M_1&=&-\frac{\kappa[(p^2-q)(\Delta_0+1)+1-q^3)}{2(p^2-q^2)}, \quad M_2=\frac{\kappa[(p^2+q)(\Delta_0-1)-1-q^3)}{2(p^2-q^2)}, \nonumber\\ J_1&=&\frac{(1+q)[(2-p^2)\Delta_0-4q+p^2q]M_1^2}{2p}, \quad J_2=\frac{(1-q)[(2-p^2)\Delta_0-4q+p^2q]M_2^2}{2p}, \nonumber\\ \label{MiJi_5} \eea
whence it follows immediately that the sources are corotating. One can also see that $M_1$ and $M_2$ always take different values, which means that the configurations (\ref{par_5}) are strictly {\it asymmetric}.

The black-hole sector of this subfamily of KC spacetimes constitute the values of the parameter $q$ from the interval $-1<q<-1/\sqrt{2}$ (the corresponding values of $p$ can have any sign), ensuring $M>0$, $M_i>0$. A peculiar feature of the latter black-hole configurations is that, for all of them, the ratio $|J|/M^2$ involving total Komar quantities is greater than 1; in other words, from the point of view of the usual single Kerr geometry, these binary systems are seen by a distant observer as a hyperextreme object. It is also curious that, whereas the lower constituent always verifies the inequality $|J_2|/M_2^2>1$, the upper constituent verifies the analogous inequality $|J_1|/M_1^2>1$ only on the interval $-1<q<q_0$, $q_0\simeq-0.842$, while on the interval $q_0<q<-1/\sqrt{2}$ the inequality $|J_1|/M_1^2<1$ holds. Another interesting physical property of the black-hole configurations of this type is that the interaction force between the components, which can be shown to have the form
\be
{\cal F}=\frac{q}{8}[p^2\Delta_0-q(2-p^2)], \label{F_5} \ee
is negative definite on the whole interval $-1<q<-1/\sqrt{2}$. This means that the spin-spin repulsive force in such binary systems exceeds the gravitational attraction, and the strut is needed in that case to impede the constituents to move away from each other! The repulsive character of the interaction is partially supported by Fig.~3 where the SLS in Fig.~3(i) may be interpreted as a perturbed common SLS before its separation into two disconnected regions, and the SLS in Fig.~3(ii) could represent a final result of the non-stationary separation process which, in addition, has led to the creation of an unstable third region endowed with a massless ring singularity.

\section{The charging generalization of KC metric}

The results of the previous section can be generalized to the electrovac case by subjecting the KC solution to the Harrison-Ernst charging transformation \cite{Har,Ern} which yields the following two complex Ernst potentials $\E$ and $\Phi$:
\be
\E=\frac{\E_0-b^2}{1-b^2\E_0}, \quad \Phi=\frac{b(\E_0-1)}{1-b^2\E_0}, \label{EF_ev} \ee
where $\E_0$ is the Ernst potential of KC solution, and $b$ is an arbitrary real constant describing the electromagnetic field in the new electrovac solution. In terms of the polynomials $A$ and $B$ defined in (\ref{KC_sol}), the above $\E$ and $\Phi$ rewrite as
\be
\E=\frac{(1-b^2)A-(1+b^2)B}{(1-b^2)A+(1+b^2)B}, \quad \Phi=-\frac{2bB}{(1-b^2)A+(1+b^2)B}. \label{EF_AB} \ee
By setting $b=0$ in (\ref{EF_ev}) or (\ref{EF_AB}), one recovers the ``seed'' KC vacuum solution.

The potential $\Phi$ is related to the electric and magnetic components of the electromagnetic 4-potential. The electric component $A_4$ is obtainable algebraically from $\Phi$ as $A_4=-{\rm Re}(\Phi)$, while the magnetic component $A_3$ can be found from the system of the first order differential equations involving the imaginary part of $\Phi$ and corresponding metric coefficients $f$, $\omega$ (see \cite{Ern} for details of Ernst's formalism).

The complex potentials (\ref{EF_ev}) define the new metric function $f$ via the formula
\be
f={\rm Re}(\E)+\Phi\bar\Phi, \label{f_ev} \ee
a bar over a symbol denoting complex conjugation. The invariance transformation (\ref{EF_ev}) does not change the form of the metric function  $\gamma_0$ of the ``seed'' solution, so that in our case the new function $\gamma$ is equal to the function $\gamma$ of KC solution from (\ref{mf_gen}). The remaining function $\omega$ of the electrovac solution (\ref{EF_ev}) can be obtained algebraically using the formula \cite{MNo}
\be
\omega=\frac{1}{(1-b^2)^2}[\omega_0(x,y)+b^4\omega_0(-x,-y)], \label{w_ev} \ee
where $\omega_0$ is the metric function $\omega$ of the ``seed'' vacuum solution.

After these preliminary remarks, below we give the final form of the electrovac metric generalizing the KC solution; similar to formulas (\ref{mf_gen}), it involves only four polynomials of Perjes' type:
\bea f&=&\frac{N}{D}, \quad e^{2\gamma}=\frac{N}{K_0^2(x^2-y^2)^4}, \quad \omega=2J_0 y -\frac{\k(1-y^2)F}{N}, \nonumber\\
N&=&\mu^2-(x^2-1)(1-y^2)\s^2, \nonumber\\ D&=&N+\mu\pi+(1-y^2)\s\tau, \nonumber\\ F&=&(x^2-1)\s\pi+\mu\tau, \nonumber\\ \mu&=&(1-b^2)[p^2(x^2-1)^2+q^2(1-y^2)^2+(\a^2-\beta^2)(x^2-y^2)^2], \nonumber\\ \s&=&2(1-b^2)[pq(x^2-y^2)+\beta(x^2+y^2)-2\a xy], \nonumber\\ \pi&=&(4/K_0)\left((1+b^2)\{K_0 [pPx(x^2+1)+qQy(y^2+1)]\right. \nonumber\\ &+&2(1-b^2)(pQ+pP\a+qQ\beta)[pqy(x^2-y^2)+\beta y(x^2+y^2)-2\a xy^2] \nonumber\\ &-&K_0(x^2-y^2)[(pQ\a-qP\beta)x+(qP\a-pQ\beta)y]\}+2(1+b^4) \nonumber\\ &\times&\left.[(p^2+\a^2-\beta^2)x^2-(q^2\a^2+p^2\beta^2)(x^2-y^2)+2(pq+\beta)(\beta x^2-\a xy)]\right), \nonumber\\ \tau&=&(4/K_0)\left((1+b^2)\{K_0 x [(qQ\a+pP\beta)(x^2-y^2)-qP(1-y^2)] \right. \nonumber\\ &+&(1-b^2)(pQ+pP\a+qQ\beta)y[(p^2-\a^2+\beta^2)(x^2-y^2)+y^2-1] \nonumber\\ &-&pQK_0y(x^2-1)\}-(1+b^4)[2p(q\a^2-q\beta^2-p\beta)(x^2-y^2) \nonumber\\ &+&\left.(pq+\beta)(1-y^2)]\right), \nonumber\\ J_0&=&-2\k(1+b^2)(pQ+pP\a+qQ\beta)/K_0, \quad K_0=(1-b^2)(p^2+\a^2-\beta^2). \label{mf_charge} \eea
It is remarkable that the structure of the metric functions $f$, $\gamma$, $\omega$ in (\ref{mf_charge}) is identical with the structure of these functions in the vacuum KC solution (\ref{mf_gen}), while the polynomials $\mu$, $\sigma$, $\pi$, $\tau$ and the constant quantities $K_0$, $J_0$ naturally generalize the respective expressions in (\ref{mf_gen}), reducing to the latter in the limit $b=0$. Mention that the charged version of the Tomimatsu-Sato $\delta=2$ field \cite{Ern4} is contained in the above formulas as the particular case $\a=\beta=Q=0$.

The explicit form of the potentials (\ref{EF_AB}) permits one to calculate some physical characteristics of the generalized KC spacetime. The resulting expressions for the total mass $M$, total angular momentum $J$, total electric and magnetic charges ${\cal Q}$ and ${\cal B}$, and for the magnetic dipole moment ${\cal M}$ are given below:
\bea
M&=&\frac{2\k(1+b^2)(pP-pQ\a+qP\beta)}{(1-b^2)(p^2+\a^2-\beta^2)}, \nonumber\\ J&=&\frac{M[(1-b^2)(pq+\beta)M+\k(1+b^2)(qQ\a+pP\beta)]}{(1+b^2)(pP-pQ\a+qP\beta)}, \nonumber\\ {\cal Q}&=&-\frac{2b}{1+b^2}M, \quad {\cal B}=-\frac{2b}{1+b^2}J_0, \quad {\cal M}={\cal Q}\frac{J}{M}, \label{MJ_tot} \quad \eea
where the non-zero magnetic charge arises due to the presence of the non-zero quantity $J_0$, and in the asymptotically flat case ($J_0=0$) this magnetic charge ${\cal B}$ vanishes.

In order to single out from the electrovac metric (\ref{mf_charge}) the subfamily of spacetimes describing two extreme KN sources separated by a massless strut, it is necessary, like in the pure vacuum case, to impose on the metric the condition of asymptotic flatness ($J_0=0$) and the axis condition ($\omega|_{x=1}=0$). Since the Harrison-Ernst transformation (\ref{EF_ev}) preserves the asymptotic flatness of the ``seed'' vacuum solution, the former condition can be satisfied in the same manner as in the case of the KC metric, namely,
\be
\a=-\frac{Q(p+q\beta)}{pP}, \label{cond_af} \ee
while the latter condition, even though it does not permit factorization which was possible in the vacuum case, still can be easily solved for the charge parameter $b$, yielding, when (\ref{cond_af}) is taken into account:
\bea
b&=&\pm\left(\frac{b_+}{b_-}\right)^{1/4}, \nonumber\\
b_\pm&=&[\beta(p^2-Q^2)-pq(Q^2\pm pP)] \nonumber\\ &\times& [\beta^2(p^2-Q^2)-pq\beta(1\pm pP+Q^2)-p^2(1\pm pP)]. \label{cond_axis} \eea

In the same manner the special configurations of extreme Kerr sources pointed out in \cite{MRu} can be generalized to the corresponding special cases of two extreme KN constituents. The first special case is defined by the parameter choice
\be
p=0, \quad q=1, \quad P=1, \quad Q=0, \label{sc1_par} \ee
which satisfies the asymptotic flatness condition $J_0=0$. Then the axis condition $\omega(x=1)=0$ can be solved for $\a$, giving
\be
\a=\pm\left[\beta(\beta+1)-\frac{2\beta}{1-b^4}\right]^{1/2}, \label{sc1_ac} \ee
and in the limit $b=0$ the above expression reduces to the respective formula from the vacuum case.

The second special choice of the parameters satisfying the condition $J_0=0$ is
\be
P=0, \quad Q=1, \quad \beta=-p/q, \label{sc2_par} \ee
and by solving the corresponding axis condition for $b$ we get
\be
b=\pm\left[\frac{(\a-p)(q^2\a^2-pq^2\a-p^2)}{(\a+p)(q^2\a^2+pq^2\a-p^2)}\right]^{1/4}. \label{sc2_b} \ee

Formulas (\ref{cond_af})-(\ref{sc2_b}) define analytically three main subfamilies of binary configurations composed of extreme KN constituents. At the same time, a non-availability of the general analytic expressions for the individual Komar quantities considerably complicates the identification and analysis of the black-hole sector of such configurations, suggesting the use of a numerical approach in the case of non-identical components. In view of that, in what follows we shall consider only two particular members of the subfamily (\ref{sc1_par}) allowing for the {\it analytical} formulas of Komar masses: the first one describing two identical corotating extreme sources, and the second one -- two counterrotating identical extreme sources. In both these cases, the mass and charge of each source are equal half the corresponding total quantity.

\subsection{Systems of two corotating extreme KN black holes}

The case of identical KN constituents with parallel angular momenta is obtainable from the general formulas by setting
\be
\a=0, \quad P=1, \quad Q=0. \label{sc_par} \ee
This choice satisfies the asymptotic flatness condition and simplifies the expression (\ref{cond_axis}) for the charge parameter $b$ fulfilling the axis condition:
\be
b=\pm\left(\frac{b_+}{b_-}\right)^{1/4} =\pm\left[\frac{(\beta-q)[p\beta^2-(q\beta+p)(1+p)]} {(\beta+q)[p\beta^2-(q\beta+p)(1-p)]}\right]^{1/4}. \label{b_cor} \ee

In this particular case the polynomials $\mu$, $\s$, $\pi$, $\tau$ in (\ref{mf_charge}) take the form
\bea \mu&=&(1-b^2)[p^2(x^2-1)^2+q^2(1-y^2)^2-\beta^2(x^2-y^2)^2], \nonumber\\ \s&=&2(1-b^2)[pq(x^2-y^2)+\beta(x^2+y^2)], \nonumber\\ \pi&=&\frac{8}{b_-(1-b^2)}\{x[2p\beta(p+q\beta)-pq-\beta] [p(x^2+1)+q\beta(x^2-y^2)] \nonumber\\ &+&2(q-p\beta)[(p+q\beta)^2x^2+p^2\beta^2y^2]\}, \nonumber\\ \tau&=&\frac{8(x-1)}{b_-(1-b^2)}\{[p\beta(x^2-y^2)-q(1-y^2)] [2p\beta(p+q\beta)-pq-\beta] \nonumber\\ &+&p\beta(pq-p^2\beta+q^2\beta)(x+1)\}, \label{mu_cor} \eea
while $K_0=(1-b^2)(p^2-\beta^2)$. Note that, for convenience, we have left the factor $(1-b^2)$, canceling out in various formulas of (\ref{mf_charge}), without substituting in it explicitly the value (\ref{b_cor}) for $b$.

For the mass, angular momentum and charge of each extreme constituent we have the expressions
\bea
M_1&=&M_2=\frac{\kappa(1+b^2)(p+q\beta)}{(1-b^2)(p^2-\beta^2)}, \nonumber\\
J_1&=&J_2=\frac{\kappa^2(1+b^2)[(2q-p\beta)(p^2+\beta^2)+4p\beta]} {(1-b^2)(p^2-\beta^2)^2}, \nonumber\\
{\cal Q}_1&=&{\cal Q}_2=-\frac{2\kappa b(p+q\beta)}{(1-b^2)(p^2-\beta^2)}, \label{Kom_cor} \eea
whence it is easy to see that the masses and charges of the constituents verify the inequality $M_i^2>{\cal Q}_i^2$.

The black-hole sector of this electrovac solution can be analyzed analytically, and during the analysis one should take into account that $b^2$ defined by (\ref{b_cor}) can be less or greater than unity. By introducing the notations
\bea
\beta_0^{(\pm)}&=&\frac{q^2-p^2}{4pq}\pm\frac14\sqrt{\frac{1+4p^2q^2}{p^2q^2}}, \quad \beta_1^{(\pm)}=\frac{q(1+p)}{2p}\pm\frac12 \sqrt{\frac{(1+p)(4p^2+pq^2+q^2)}{p^2}}, \nonumber \\ \beta_2^{(\pm)}&=&\frac{q(1-p)}{2p}\pm\frac12 \sqrt{\frac{(1+p)(4p^2-pq^2+q^2)}{p^2}}, \quad q_0=\sqrt{\frac{1}{2}(\sqrt{5}-1)},
\label{bh_cor} \eea
the parameter ranges leading to the positive values of the mass have been summarized in two tables. In Table~I we give the intervals of the parameters $q$ and $\beta$ (together with the choice of sign of $p$) defining the black-hole configurations with $b^2<1$. Table~II deals with the black-hole sector when $b^2>1$.

The main distinction of the configurations consisting of corotating identical KN extreme black holes from the analogous configurations composed of the extreme Kerr black holes which were considered in \cite{MRu,CHR} is the following: whereas for the latter configurations the inequality $1<|J_i|/M_i^2<2$ for the individual Komar quantities holds in general, the former binary systems, due to the presence of a non-zero electric charge, do not necessarily satisfy the above inequality, so that the ratio $|J_i|/M_i^2$ for the KN constituents can be not only greater but also less than 1. In Fig.~4(i) we have plotted the SLS of electrovac configuration defined by the parameter values
\be
\kappa=2, \quad q\simeq0.954, \quad p=-0.3, \quad \beta=-2, \quad b\simeq0.874. \label{pc_5} \ee
The corresponding physical characteristics of interest are
\be
M_i\simeq 8.454, \quad \frac{J_i}{M_i^2}\simeq 0.212, \quad \frac{{\cal Q}_i}{M_i}\simeq -0.991, \label{MJQ_5} \ee
whence it follows that the ratio $J_i/M_i^2$ is much less than 1 due to a large value of the electric charge which is almost equal in absolute value to the mass. Nonetheless, a curious fact is that in this particular example we still do have a quantity which is greater than 1, although in the case of a single KN extreme black hole the same quantity is known to be equal to unity:
\be
\frac{J_i^2}{M_i^4}+\frac{{\cal Q}_i^2}{M_i^2}\simeq 1.027. \label{JQ_5} \ee
It is worth mentioning that for the total characteristics $M$, $J$ and ${\cal Q}$ the above quantity is approximately equal to 0.993, so that a distant observer would see this specific binary configuration as a subextreme KN black hole.

In Fig.~4(ii) we have plotted the SLS of another particular configuration, defined by
\be
\kappa=2, \quad q=0.75, \quad p\simeq-0.661, \quad \beta=0.8, \quad b\simeq0.448, \label{pc_6} \ee
which clearly demonstrates the new distinguishing features introduced by the electromagnetic field. Thus, for the parameter choice (\ref{pc_6}) we have
\be
M_i\simeq 0.912, \quad \frac{J_i}{M_i^2}\simeq 12.303, \quad \frac{{\cal Q}_i}{M_i}\simeq -0.747, \label{MJQ_6} \ee
and the above ratio $J_i/M_i^2$ considerably exceeds the maximum value 2 that two identical corotating Kerr black holes may have! This can be explained by the presence of a massless ring singularity in the equatorial plane, absent in the vacuum case, which can be associated with an external source of the electromagnetic field.

Let us observe for completeness that, unlike in the pure vacuum case, the strut between the corotating KN extreme constituents can be removed even when the latter have positive Komar masses. Since the Harrison-Ernst transformation does not change the form of the metric function $\gamma$, the equilibrium condition is the same as in the vacuum case, i.e., $p^2=q^2=1/2$ (cf. Ref.~\cite{MRu}). Then it is not difficult to find the parameter ranges at which equilibrium occurs and the masses take positive values:
\be
p=q=\pm\frac{1}{\sqrt{2}}, \quad -1<\beta<-\frac{1}{\sqrt{2}}, \quad b^2=\left(\frac{\beta-1\mp\sqrt{2}}{\beta-1\pm\sqrt{2}}\right)^{1/2}, \label{eq_1} \ee
and
\be
p=-q=\pm\frac{1}{\sqrt{2}}, \quad \frac{1}{\sqrt{2}}<\beta<1, \quad b^2=\left(\frac{\beta+1\pm\sqrt{2}}{\beta+1\mp\sqrt{2}}\right)^{1/2}. \label{eq_2} \ee

The equilibrium configurations of that kind were studied numerically in the paper \cite{MSM} where it was shown that they are always accompanied by a massless ring singularity lying in the equatorial plane (like the one in Fig.~4(ii)) which could represent some source of the electromagnetic field.

\subsection{Systems of two counterrotating extreme KN black holes}

We now turn to the second particular case of the electrovac metric (\ref{mf_charge}) arising from the general formulas by choosing the parameters in the form
\be
p=1, \quad q=0, \quad \a=-Q/P, \quad \beta=0. \label{par_as} \ee

The above choice of the parameters satisfies both the asymptotic flatness and axis conditions, independently of the value of the charge parameter $b$. The metric functions in this case assume a remarkably simple form, and we write them down below together with the corresponding Ernst potentials:
\bea \E&=&\frac{(1-b^2)A-(1+b^2)B}{(1-b^2)A+(1+b^2)B}, \quad \Phi=-\frac{2bB}{(1-b^2)A+(1+b^2)B}, \nonumber\\ f&=&\frac{(1-b^2)^2N}{D}, \quad e^{2\gamma}=\frac{N}{(x^2-y^2)^4}, \quad \omega=-\frac{4\kappa P^2Q y(x^2-1)(1-y^2)F}{(1-b^2)^2N}, \nonumber\\ A&=&P^2(x^4-1)+Q^2(x^2-y^2)^2+2iPQ(x^2+y^2-2x^2y^2), \nonumber\\ B&=&2(P-iQ)Px[P(x^2-1)+iQ(x^2-y^2)], \nonumber\\ N&=&[P^2(x^2-1)^2+Q^2(x^2-y^2)^2]^2-16P^2Q^2x^2y^2(x^2-1)(1-y^2), \nonumber\\ D&=&\{(1-b^2)[P^2(x^4-1)+Q^2(x^2-y^2)^2]+2(1+b^2)Px[P^2(x^2-1)+Q^2(x^2-y^2)]\}^2 \nonumber\\ &+&4P^2Q^2[(1-b^2)(x^2+y^2-2x^2y^2)+(1+b^2)Px(1-y^2)]^2, \nonumber\\ F&=&(1-b^4)[Q^2(x^2-y^2)(3x^2+y^2)+P^2(3x^4+6x^2-1)]+8(1+b^4)Px^3. \label{mf_as} \eea

Formulas (\ref{mf_as}) describe configurations of two KN extreme sources that are identical and counterrotating. In the vacuum limit ($b=0$) the metric (\ref{mf_as}) reduces to the vacuum spacetime for two counterrotating  extreme Kerr sources which was considered in the paper \cite{MRRS}. Note that the above metric is equatorially antisymmetric \cite{EMR} as its coefficients transform like $f\to f$, $\gamma\to\gamma$, $\omega\to-\omega$ under the substitution $y\to-y$.

The establishing of the black-hole sector of this binary configuration does not represent any difficulty. Indeed, the individual masses and charges of the constituents have the form
\be
M_1=M_2=\frac{\kappa P(1+b^2)}{1-b^2}, \quad {\cal Q}_1={\cal Q}_2=-\frac{2\kappa b P}{1-b^2}, \label{MiQi_as} \ee
hence the masses take positive values when $P>0$, $b^2<1$, and $P<0$, $b^2>1$. Since the total angular momentum of this configuration is equal to zero, the individual angular momenta are such that $J_1=-J_2$. Finding the explicit form of $J_1$ and $J_2$ is, however, a difficult and very laborious technical problem whose resolution requires a preliminary construction of the individual angular momenta for the Bret\'on-Manko solution \cite{BMa} describing two counterrotating charged particles, and then taking appropriately the extreme limit. After performing all the necessary calculations, we have been able to arrive eventually at the following remarkably simple final result:
\be
J_1=-J_2=\frac{M_1\kappa P[1-b^4+P(1+b^4)]}{Q(1-b^4)}, \label{Ji_as} \ee
which permits us to verify analytically whether each of the two counterrotating identical KN extreme black holes satisfies the same relation between its mass, angular momentum and charge as in the case of a single extreme KN black hole. For this purpose we find from (\ref{MiQi_as}) and (\ref{Ji_as}) that
\be
M_i^2-\frac{J_i^2}{M_i^2}-Q_i^2=-\frac{2\kappa^2 P^3[1-b^8+P(1+b^8)]}{Q^2(1-b^4)^2}, \label{MJQ_as} \ee
whence it follows immediately that the right-hand side of (\ref{MJQ_as}) is a strictly {\it negative} quantity for all $P$ and $b$ defining the black-hole configurations, i.e., for $P>0$, $b^2<1$ and $P<0$, $b^2>1$. Taking into account that for a single KN extreme black hole the right-hand side of (\ref{MJQ_as}) must be equal to {\it zero}, we conclude that the constituents in all black-hole configurations described by the metric (\ref{mf_as}) would look, from the point of view of the usual single KN solution, as hyperextreme objects.

The presence of the electric charge in the black-hole configurations of the solution (\ref{mf_as}) enables the ratio $|J_i|/M_i^2$ to be less, equal or greater than~1, which is in analogy with the case of corotating identical black holes considered in the previous subsection. An example of the values of the parameters leading to $|J_i|/M_i^2<1$ is the following:
\be
\frac{1}{\sqrt{5}}<Q<1, \quad P>0, \quad \frac{\sqrt{Q^2+2PQ}-Q}{1-P+Q}<b^2<1. \label{rat_as} \ee

In Fig.~5 we have plotted the SLS of two particular ``antisymmetric'' black-hole configurations defined by
\be
\kappa=2, \quad P=Q=\frac{1}{\sqrt{2}}, \quad b=\frac{1}{2}, \label{pc_7} \ee
and
\be
\kappa=2, \quad P=-\frac{24}{25}, \quad Q=\frac{7}{25}, \quad b=-\frac{3}{2}, \label{pc_8} \ee
in both these cases the corresponding spacetimes being completely regular outside the black hole horizons defined by the points $x=1$, $y=1$ and $x=1$, $y=-1$ of the symmetry axis.

Although the value of the charge parameter $b$ does not affect the form of SLS which are determined exclusively by the values of the parameters $\kappa$, $P$ and $Q$ ($\kappa$ being a sort of a scale parameter), the constant $b$ is able to affect the singularity structure of the electrovac solution (compared with the analogous structure of the vacuum ``seed'' solution) which is defined by zeros of the denominator of the potential $\E$. For instance, the spacetimes of the above electrovac configuration (\ref{pc_7}) and of the corresponding vacuum configuration with $b=0$ are both regular outside the horizons; on the other hand, the electrovac configuration (\ref{pc_8}) is regular everywhere outside the horizons, whereas its vacuum ``seed'' has two massless ring singularities off the symmetry axis due to the negative mass of the constituents in the latter solution.

In general, metric (\ref{mf_as}) illustrates well a remarkable geometric property of the Harrison-Ernst charging transformation: {\it this transformation establishes correspondence between the vacuum and electrovacuum solutions possessing the same stationary limit surface, even though the singularity structures of the solutions can be different.}

\section{Discussion and conclusion}

In view of of the recent fundamental findings on the relation between the stationary axisymmetric and non-stationary axisymmetric systems \cite{DOr} that have led to the development of powerful numerical methods permitting one to analyze various non-stationary models of multiple black holes, the study of stationary axisymmetric binary black-hole configurations acquires ever greater importance as an indispensable ingredient of that more general study, providing one with a reliable physical information obtainable from exact solutions. In the present paper we have obtained various generic analytical results for the systems of two interacting Kerr or KN extreme black holes which considerably simplify the future use of such configurations in the basic research and applications. Although the two-black-hole solutions considered by us are endowed with a conical singularity in general, the majority of them are completely regular everywhere outside the black-hole horizons ($x=1$, $y=\pm1$). In this respect, the issue of a massless ring singularity off the symmetry axis emerging in certain binary black-hole systems is intriguing and needs further investigation. Before the paper \cite{MRu}, such naked singularities in the framework of the double-Kerr solution have been always associated exclusively with a {\it negative} mass of one of the constituents, and their presence in the configurations involving negative mass was enough for labeling those configurations as pathological and unphysical. However, the massless ring singularities in the binary black-hole systems with {\it positive} Komar masses of both constituents obviously need a different, probably more physical, explanation of their origin. The one we have proposed in this paper relies on the fact that these novel ring singularities seem to arise only in the configurations with a common SLS, the one which has not yet been completely formed or is going to be disintegrated, like in Figs.~2(ii) and 3(i), or with a SLS just disintegrated into more than two separated regions, like in Fig.~3(ii). It seems plausible to suppose that the processes of the formation and disintegration of a common SLS, once initiated, may have a very violent, non-stationary and turbulent character, even when the two black holes have already stopped their mutual movement thanks, say, to a strut. Then a massless ring singularity is needed to stop artificially the non-stationary processes related to the SLS, and as such it plays the same role with respect to the SLS as the conical singularity plays with respect to the black holes -- keeping up the stationarity of the configuration. At the same time, we do not exclude that this singularity could only appear in the stationary black-hole systems, and that it does not arise for instance in the non-stationary axisymmetric configurations where there is no need to control the evolution of the SLS, unless the axisymmetry of the system is affected. It is probably worth mentioning that if one does not want to be involved in the problem of singularities, he must give preference to the systems of counterrotating  black holes because, as was observed already by Oohara and Sato \cite{OSa} in the context of the double-Kerr solution, such configurations do not form mergers of SLS which are a requisite for the formation of massless ring singularities in the case of extreme Kerr black holes.

The singularity discussed above should not be confused with the ring singularity of the electrovac configuration of KN black holes shown in Fig.~4(ii). Although the latter singularity is also massless, it has the electromagnetic nature and hence has no analogs in the pure vacuum case. We would like to emphasize that the metric (\ref{mf_charge}) is not the most general electrovac generalization of the KC solution, so it is quite possible that there could exist other configurations of extreme KN black holes apart from those already considered here. The task of completing the list of the binary systems describing the KN extreme black holes is very difficult from the technical point of view as its implementation requires the use of the 9-parameter electrovac metric constructed in \cite{MSM}. Nevertheless we do not lose hopes that such a task could be performed by someone in the future.

\section*{Acknowledgements}
We are grateful to the referee for valuable remarks and suggestions. This work was partially supported by CONACyT, Mexico, and by MCyT of Spain under the Project FIS2009-07238.

\newpage

\begin{table}[htb]
\caption{The black-hole sector of the parameters $q$, $p$ and $\beta$ in the case $b^2<1$. To choose a concrete parameter set $\{q,p,\beta\}$, one has first choose the value of $q$ from one of the intervals given in the table, then find the corresponding value of $p=\pm\sqrt{1-q^2}$, choosing for it an appropriate sign according to the table, and finally assign to $\beta$ some value from the interval defined by the selected values of $q$ and $p$.}
\begin{center}
\begin{tabular}{lll}
\hline \hline $q$  & $p$
 & $\beta$ \\ \hline $(-1,-q_0)$ \hspace{2cm} & $<0$, $>0$  \hspace{2cm} & $(\beta_0^{(+)},\beta_1^{(+)})$ \\
$(-q_0,-\frac{1}{\sqrt{2}})$  & $<0$, $>0$  & $(\beta_0^{(+)},\beta_1^{(+)})$  \\
$(-q_0,-\frac{1}{\sqrt{2}})$  & $<0$  & $(-\frac{p}{q},q)$ \\
$(-\frac{1}{\sqrt{2}},0)$  & $<0$ & $(-\frac{p}{q},\beta_0^{(-)})$ \\
$(-\frac{1}{\sqrt{2}},0)$  & $>0$ & $(q,\beta_0^{(-)})$ \\
$(0,\frac{1}{\sqrt{2}})$  & $<0$ & $(\beta_0^{(+)},-\frac{p}{q})$ \\
$(0,\frac{1}{\sqrt{2}})$  & $>0$ & $(\beta_0^{(+)},q)$ \\
$(\frac{1}{\sqrt{2}},q_0)$  & $<0$, $>0$  & $(\beta_1^{(-)},\beta_0^{(-)})$ \\
$(\frac{1}{\sqrt{2}},q_0)$  & $<0$ & $(q,-\frac{p}{q})$ \\
$(q_0,1)$  & $<0$, $>0$  & $(\beta_1^{(-)},\beta_0^{(-)})$ \\ \hline \hline
\end{tabular}
\end{center}
\end{table}

\newpage

\begin{table}[htb]
\caption{The black-hole sector of the parameters $q$, $p$ and $\beta$ in the case $b^2>1$.}
\begin{center}
\begin{tabular}{lll}
\hline \hline $q$  & $p$
 & $\beta$ \\ \hline $(-1,-q_0)$ \hspace{2cm} & $<0$, $>0$  \hspace{2cm} & $(\beta_2^{(-)},\beta_0^{(-)})$ \\
$(-q_0,-\frac{1}{\sqrt{2}})$  & $<0$, $>0$  & $(\beta_2^{(-)},\beta_0^{(-)})$  \\
$(-q_0,-\frac{1}{\sqrt{2}})$  & $>0$  & $(-q,-\frac{p}{q})$ \\
$(-\frac{1}{\sqrt{2}},0)$  & $<0$ & $(\beta_0^{(+)},-q)$ \\
$(-\frac{1}{\sqrt{2}},0)$  & $>0$ & $(\beta_0^{(+)},-\frac{p}{q})$ \\
$(0,\frac{1}{\sqrt{2}})$  & $<0$ & $(-q,\beta_0^{(-)})$ \\
$(0,\frac{1}{\sqrt{2}})$  & $>0$ & $(-\frac{p}{q},\beta_0^{(-)})$ \\
$(\frac{1}{\sqrt{2}},q_0)$  & $<0$, $>0$  & $(\beta_0^{(+)},\beta_2^{(+)})$ \\
$(\frac{1}{\sqrt{2}},q_0)$  & $>0$ & $(-\frac{p}{q},-q)$ \\
$(q_0,1)$  & $<0$, $>0$  & $(\beta_0^{(+)},\beta_2^{(+)})$ \\ \hline \hline
\end{tabular}
\end{center}
\end{table}

\newpage

\begin{figure}[htb]
\centerline{\epsfysize=70mm\epsffile{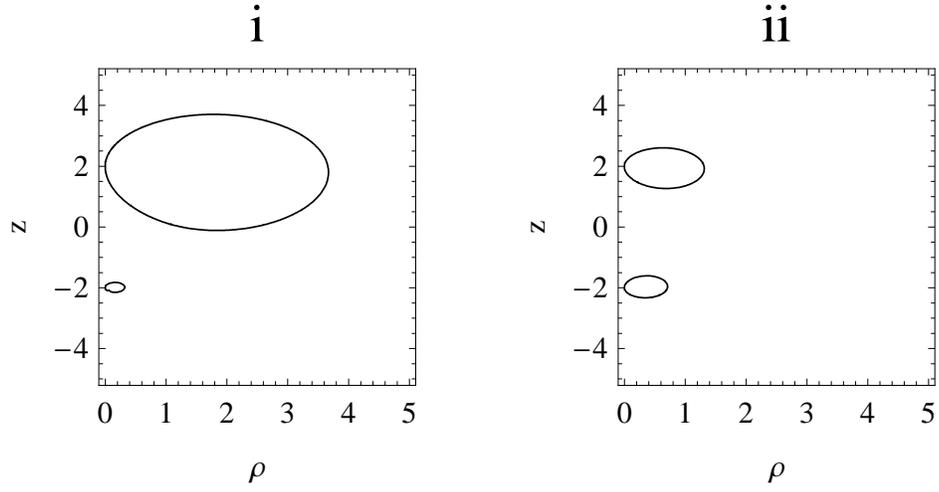}} \caption{Typical shape of SLS in the case of two counterrotating extreme non-identical Kerr black holes.}
\end{figure}

\begin{figure}[htb]
\centerline{\epsfysize=70mm\epsffile{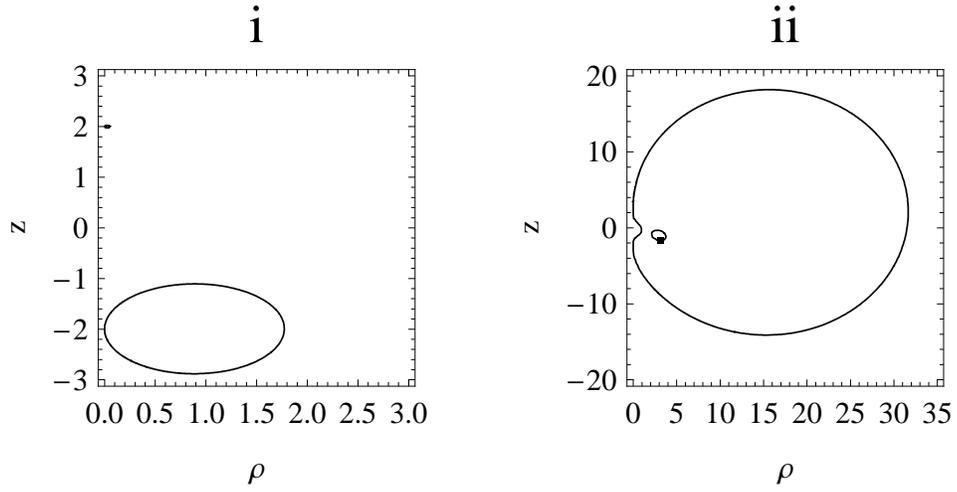}} \caption{Two examples of SLS in the binary configurations of extreme corotating Kerr black holes: ($i$) the case of a SLS with two disconnected regions, not leading to the formation of a naked singularity off the symmetry axis; ($ii$) the case of a common SLS with an internal region characterized by a massless ring singularity.}
\end{figure}

\begin{figure}[htb]
\centerline{\epsfysize=70mm\epsffile{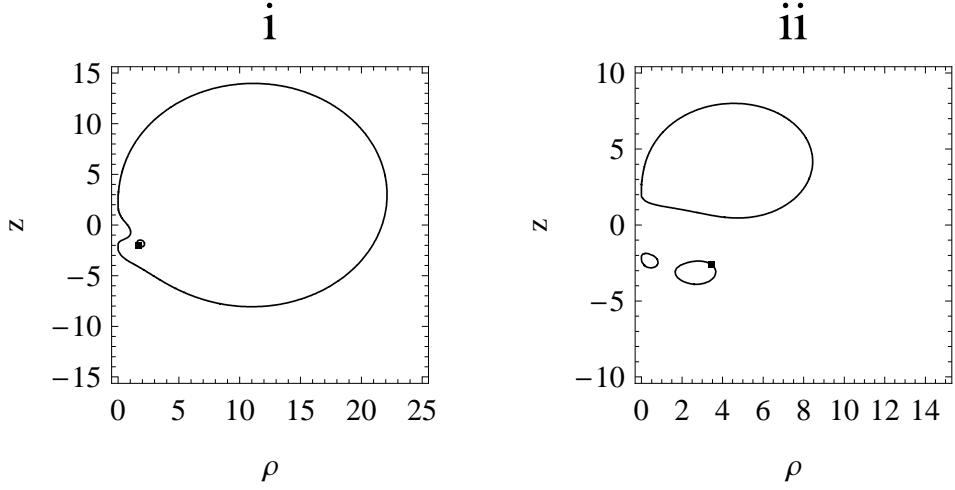}} \caption{The SLS in the special corotating case defined by (\ref{par_5}): ($i$) the extreme Kerr black holes with a common SLS and a massless naked singularity (the parameters are assigned the following values: $\kappa=2, p=3/5, q=-4/5$); ($ii$) formation of two disconnected regions accompanied by the third one with a massless ring singularity, the values of the parameters are $\kappa=2, p=7/25, q=-24/25$.}
\end{figure}

\begin{figure}[htb]
\centerline{\epsfysize=70mm\epsffile{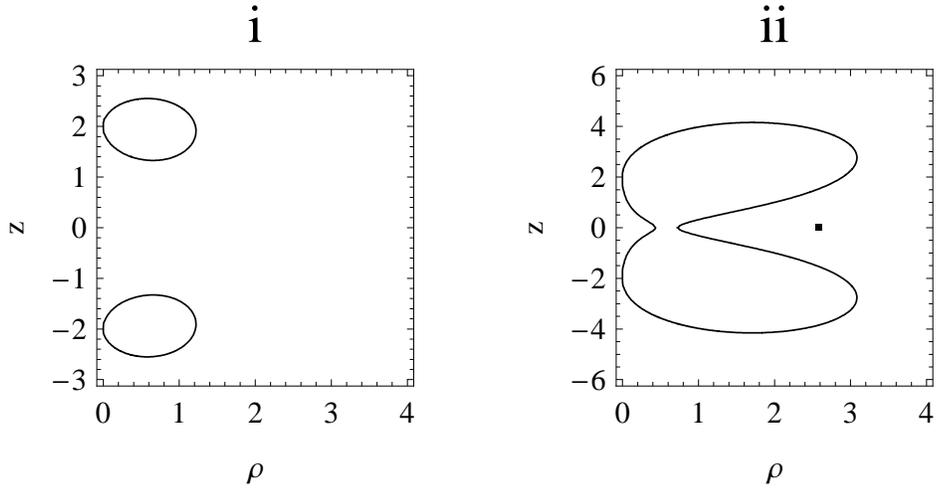}} \caption{The SLS of two identical corotating extreme KN black holes: ($i$) SLS with two disconnected regions and without naked singularities off the axis; ($ii$) the case of a common SLS and a naked ring singularity in the equatorial plane.}
\end{figure}

\begin{figure}[htb]
\centerline{\epsfysize=70mm\epsffile{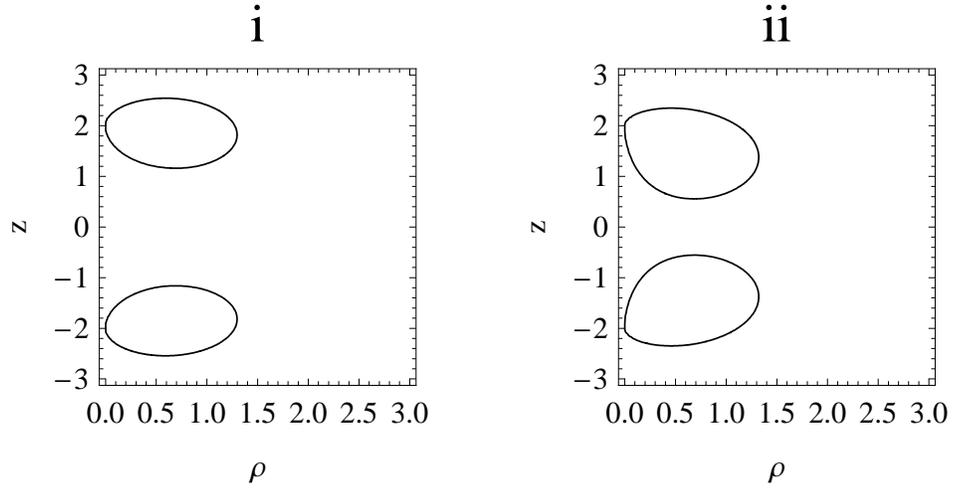}} \caption{The SLS of two identical counterrotating extreme KN black holes corresponding to the choice of the parameters (\ref{pc_7}) and (\ref{pc_8}).}
\end{figure}

\end{document}